\documentclass[a4paper,12pt]{article}

\usepackage{ifpdf}

\newif\ifpdf
\ifx\pdfoutput\undefined
  \pdffalse
\else
  \pdfoutput=1
  \pdftrue
\fi

\RequirePackage{xspace} %
\RequirePackage{subfigure} %
\RequirePackage[centertags]{amsmath} %
\RequirePackage{amssymb}
\RequirePackage{wrapfig} %
\RequirePackage{calc} %
\RequirePackage{ifthen}
\RequirePackage{tabularx} %
\RequirePackage{flafter} %
\RequirePackage{fancyhdr} %

\ifpdf
  \RequirePackage[pdftex]{color}%
  \RequirePackage{colortbl}%
  \RequirePackage{array}%
  \RequirePackage[pdftex]{graphicx}

  \RequirePackage[ pdftex, plainpages = false, pdfpagelabels,
                 pdfpagelayout = useoutlines,
                 bookmarks,
                 breaklinks = true,
                 linktocpage,
                 colorlinks = true,
                 linkcolor = blue,
                 urlcolor  = blue,
                 citecolor = blue,
                 anchorcolor = blue,
                 hyperindex = true,
                 hyperfigures
                 ]{hyperref}

\else
  \RequirePackage{color}
  \RequirePackage{colortbl}
   \RequirePackage{array}
  \RequirePackage[dvips]{graphicx}
  \RequirePackage{hyperref}
  \usepackage{rotating}
\fi

\usepackage{makeidx} 
\usepackage{setspace} 
\usepackage{rotating} 
\usepackage{ecltree}
\usepackage{epic}
\usepackage{supertabular}  
\usepackage{color}
\usepackage{exscale}
\usepackage{fontenc}
\usepackage{ifthen}
\usepackage{latexsym}
\usepackage{makeidx}
\usepackage{syntonly}
\usepackage{inputenc}
\usepackage{graphicx}
\usepackage{setspace}
\usepackage{caption2}
\usepackage[english]{babel}
\usepackage[square, comma,numbers,sort&compress]{natbib}
\usepackage{hypernat}
\usepackage{boxedminipage}
\usepackage{framed}
\usepackage{longtable}
\usepackage[all]{hypcap}    
\usepackage{algorithm2e}
\usepackage{algorithmic}
\usepackage{lscape}
\usepackage{pdflscape}
\usepackage[T1]{fontenc}
\usepackage{microtype}

\newcommand{\note}[1]{\marginpar[left]{\singlespace \tiny #1}}

\renewcommand{\sectionmark}[1]%
      {\markright{\thesection\ #1}} 

\renewcommand{\note}[1]{}

\doublespace 

\DisableLigatures[f]{encoding = *, family = * }

\begin{document}
\begin{center}
{\Large Using the stress function in the flow of generalized Newtonian fluids through conduits with
non-circular or multiply connected cross sections}
\par\end{center}{\Large \par}

\begin{center}
Taha Sochi
\par\end{center}

\begin{center}
{\scriptsize University College London, Department of Physics \& Astronomy, Gower Street, London,
WC1E 6BT \\ Email: t.sochi@ucl.ac.uk.}
\par\end{center}

\begin{abstract}
\noindent We investigate the possibility that the spatial dependency of stress in generalized
Newtonian flow systems is a function of the applied pressure field and the conduit geometry but not
of the fluid rheology. This possibility is well established for the case of a one-dimensional flow
through simply connected regions, specifically tubes of circular uniform cross sections and plane
thin slits. If it can also be established for the more general case of generalized Newtonian flow
through non-circular or multiply connected geometries, such as the two-dimensional flow through
conduits of rectangular or elliptical cross sections or the flow through annular circular pipes,
then analytical or semi-analytical or highly accurate numerical solutions; regarding stress, rate
of strain, velocity profile and volumetric flow rate; for these geometries can be obtained from the
stress function, which can be easily obtained from the Newtonian case, in combination with the
constitutive rheological relation for the particular non-Newtonian fluid, as done previously for
the case of the one-dimensional flow through simply connected regions.

\vspace{0.3cm}


\par\end{abstract}

\clearpage
\section{Introduction} \label{Introduction}

The flow of generalized Newtonian fluids through conduits with circular and non-circular simply
connected cross sections, such as those of elliptical or rectangular or triangular shape, and
multiply connected cross sections like circular annulus, is commonplace in many biological systems
and technological applications such as the transport of biological fluids in living organisms, the
shipping of industrial liquids and the distribution of coolants in temperature regulating devices.

There are many investigations related to the flow of Newtonian fluids through conduits with
non-circular cross sections or with multiple connectivity (e.g. \cite{Claiborne1951, Sparrow1962,
Sastry1964, RatkowskyE1968, ShahL1971, Richardson1980, YovanovichM1997, Lekner2007, Lekner2009,
GeorgiouK2013, KaoullasG2013, WuHL2013, WiwatanapatapheeWS2014}), and less on the flow of
non-Newtonian fluids through such conduits (e.g. \cite{Schechter1961, Middleman1965, KozickiCT1966,
Miller1972, EscudierOPS2002, MuzychkaE2008}). Several methods have been used in these
investigations such as direct application of Laplace and Poisson equations, complex analysis,
conformal mapping, variational methods, and numerical discretization techniques
\cite{Schechter1961, Sastry1964, Middleman1965, ShahL1971, YovanovichM1997,
PapanastasiouGABook1999, Lekner2007} as well as experimental examination \cite{EscudierOPS2002}.

In this paper we investigate the possibility that the stress function for generalized Newtonian
fluids in multi-dimensional and multiple connectivity flow is universal, i.e. it is the same for
Newtonian and non-Newtonian fluids, and hence the flow fields of generalized Newtonian fluids of
non-Newtonian rheology through conduits with non-circular or multiply connected cross sections can
be obtained by acquiring the stress, as a function of the spatial coordinates of the conduit cross
section, from the Newtonian case. The stress function can then be utilized in combination with the
rheological constitutive relation of the particular non-Newtonian fluid to obtain the flow field
parameters which include the shear rate, as a function of the spatial coordinates of the cross
section, and thereby the flow velocity profile and subsequently the volumetric flow rate. If this
method is established, through the establishment of the universality of the stress function, it
will be simple, general, reliable and easy to implement; moreover it can produce highly accurate
solutions for the flow of generalized Newtonian fluids of non-Newtonian rheology in those
geometries.

The plan for this paper is that in section \ref{Method} we explain the method and the supporting
argument for the universality of the stress function in general terms stating the relevant
assumptions and restrictions. This will be followed in section \ref{Examples} by a few examples of
the stress function for conduits of non-circular or multiply connected cross sections which are
obtained from the Newtonian flow case. The study will be concluded in section \ref{Conclusions}
with general briefing and discussion.

\section{Method}\label{Method}

Here, we assume a laminar, incompressible, steady state, rectilinear, isothermal, pressure-driven,
fully-developed, creeping flow of a purely-viscous, time-independent generalized Newtonian fluid
and hence history-dependent fluids, like viscoelastic and thixotropic, are excluded. We also
exclude viscoplastic fluids, even if they are classified as generalized Newtonian fluids, due to
the complications introduced by the presence of yield stress and the failure of the available
viscoplastic models to account for these complications. The effect of any potential secondary flow
under these conditions is negligible.

The effects of external body forces, such as gravity, as well as the edge effects at the entry and
exit zones of the conduit are assumed insignificant. Dependencies on physical factors like
temperature, which are not related to deformation, are also ignored assuming fixed conditions or
negligible contribution from these factors. The flow is also assumed to be in shear mode with no
significant extensional contributions. Moreover, the pressure is assumed to be a sole function of
the axial dimension in the flow direction.

Concerning the type of conduit, we consider cylindrical ducts of uniform cross sections (i.e.
having constant shape and size in the flow axial direction) with non-circular simply or/and
multiply connected cross section geometry. Rigid mechanical properties of the conduit wall are
assumed and hence the conduit wall is not deformable under the considered range of pressure. As for
the boundary conditions, no-slip at the conduit wall is assumed and hence a zero velocity condition
at the fluid-solid interface is maintained.

We start from the momentum equation for the fluid flow which is given by \cite{BirdbookAH1987,
BirdSLbook2002}

\begin{equation}\label{MomEq1}
\rho\frac{D\mathbf{v}}{Dt}=-\nabla p-\nabla.\boldsymbol{\tau}+\rho\mathbf{g}
\end{equation}
where $\rho$ is the fluid mass density, $\frac{D}{Dt}$ is the material derivative, $\mathbf{v}$ is
the fluid velocity vector, $t$ is the time, $\nabla$ is the gradient operator, $p$ is the pressure,
$\boldsymbol{\tau}$ is the deviatoric or extra stress tensor, and $\mathbf{g}$ is the gravitational
acceleration vector. Now, for a steady state creeping flow with negligible body forces we can
neglect the time rate, convection and gravitational terms, and hence Equation \ref{MomEq1} becomes

\begin{equation}\label{MomEq2}
\nabla.\boldsymbol{\tau}=-\nabla p
\end{equation}
This equation can be simplified for two dimensional flow in a Cartesian coordinates system, where
the pressure gradient is in the stress-invariant $z$ direction, into the following form of the
$z$-component

\begin{equation}\label{MomEq3}
\frac{\partial\tau_{xz}}{\partial x}+\frac{\partial\tau_{yz}}{\partial y}=-\frac{\partial
p}{\partial z}
\end{equation}
The stress in the last equation is dependent on the conduit spatial dimensions (geometry) and the
applied pressure but is independent of the fluid rheology since the equation does not contain any
rheological parameter. The independence of the stress function from the fluid rheology is clearly
demonstrated by the absence of such dependency in the expressions of the stress function in the
Newtonian flow case; examples of which are given in section \ref{Examples}.

For a given set of stress boundary conditions the solution of this equation should be unique. All
we need to establish then is that the stress boundary conditions for generalized Newtonian fluids
of non-Newtonian rheology are the same as those for the Newtonian rheology to conclude that the
stress function (i.e. spatial dependency of stress over the conduit cross sectional region under
steady flow conditions) for non-Newtonian rheology is the same as for Newtonian rheology. Now, the
indifference of the stress boundary conditions between the non-Newtonian fluids and Newtonian
fluids is well established for the case of one-dimensional simply connected flow which is
represented by the case of circular cylindrical tubes and thin plane slits, as discussed in
\cite{SochiStress1D2015}. If the stress boundary conditions should depend on the rheology, the
indifference cannot be established even in the case of one-dimensional simply connected flow.
Hence, since we cannot see any particular consideration that can justify the restriction of such an
indifference to the one-dimensional simply connected flow, we can assume that the indifference in
the stress boundary conditions, and hence in the stress function as a whole, is applicable in
general to multi-dimensional and multiply connected flows of the assumed type.

If the assumption that the stress function is the same for the Newtonian and non-Newtonian fluids
is established, we can then obtain the stress function from the Newtonian flow case and use it for
the non-Newtonian flow cases. In most circumstances, the stress function for the Newtonian flow is
easily obtained analytically or semi-analytically (e.g. from infinite series solutions). As soon as
the stress, $\tau$, as a function of the spatial coordinates of the cross section is obtained, the
rate of strain, $\gamma$, as a function of the spatial coordinates can be easily obtained from the
fluid rheological constitutive relation which correlates the rate of shear strain to the shear
stress as long as the relation can be put in the form $\gamma=\gamma(\tau)$ where the dependency of
$\gamma$ on $\tau$ can be explicit or implicit. If $\gamma$ is an explicit function of $\tau$, as
it is the case for example in the Ellis fluid (refer to Table \ref{GTable}), then $\gamma$ can be
obtained directly by a simple substitution in the rheological relation. If, on the other hand,
$\gamma$ is an implicit function of $\tau$, as it is the case for example in the Cross fluid (refer
to Table \ref{GTable}), then $\gamma$ can be obtained numerically using a simple numerical solver
based for instance on a bisection method. In both cases, the obtained rate of strain as a function
of the spatial coordinates can be used to obtain the fluid velocity profile and subsequently the
volumetric flow rate by consecutive integrations, as detailed in \cite{SochiStress1D2015}.

In Table \ref{GTable} we present a few examples of the rheological constitutive relations for fluid
models that can be used in conjunction with the stress function to obtain the flow field
parameters. As indicated, for Carreau and Cross models, $\gamma$ is given as an implicit function
of $\tau$ and hence a simple numerical solver like bisection is required to obtain $\gamma$ as a
function of $\tau$ and hence as a function of the spatial coordinates of the conduit cross section.
In the following section, we present some examples of analytical and semi-analytical stress
functions obtained from the Newtonian flow case for some types of conduit geometry with
non-circular or multiply connected cross sections.

\begin{table} [!h]
\caption{The rate of shear strain, $\gamma$, as a function of shear stress, $\tau$, for a sample of
five non-Newtonian fluids \cite{Skellandbook1967, BirdbookAH1987, CarreaubookKC1997,
Tannerbook2000, OwensbookP2002} that can be employed in the investigated stress function approach
for generalized Newtonian flow in conduits with non-circular or multiply connected cross sectional
geometries. For Carreau and Cross models, $\gamma$ is given as an implicit function of $\tau$. The
meanings of the symbols are given in Nomenclature \S\ \ref{Nomenclature}. \label{GTable}}
\begin{center} 
{
\begin{tabular}{|l|l|}
\hline Model & Rate of Shear Strain\tabularnewline
 \hline
Power Law & $\gamma=\sqrt[n]{\frac{\tau}{k}}$\tabularnewline
Ellis &
$\gamma=\frac{\tau}{\mu_{e}}\left[1+\left(\frac{\tau}{\tau_{h}}\right)^{\alpha-1}\right]$\tabularnewline
Ree-Eyring &
$\gamma=\frac{\tau_{c}}{\mu_{r}}\sinh\left(\frac{\tau}{\tau_{c}}\right)$\tabularnewline
Carreau &
$\gamma\left[\mu_{i}+\left(\mu_{0}-\mu_{i}\right)\left(1+\lambda^{2}\gamma^{2}\right)^{\left(n-1\right)/2}\right]=\tau$
\tabularnewline
Cross &
$\gamma\left[\mu_{i}+\frac{\mu_{0}-\mu_{i}}{1+\lambda^{m}\gamma^{m}}\right]=\tau$\tabularnewline
 \hline
\end{tabular}
}
\end{center}
\end{table}

\section{Examples of Stress Functions for non-Circular and Multiply Connected Cross
Sections}\label{Examples}

Here, we present some examples of non-circular and multiply connected cross sectional shapes of the
flow conduits whose stress function can be obtained from their Newtonian flow case in the form of
analytical or infinite series solutions (refer for instance to \cite{ShahL1971, ShahLBook1978,
Whitebook1991, PapanastasiouGABook1999, WhiteBook2002, Lekner2007}). Similar expressions of stress
functions related to other cross sectional geometries, which can be derived from the Newtonian flow
case, can also be obtained from references like the above.

\begin{figure}[!h]
\centering{}
\includegraphics
[scale=0.6] {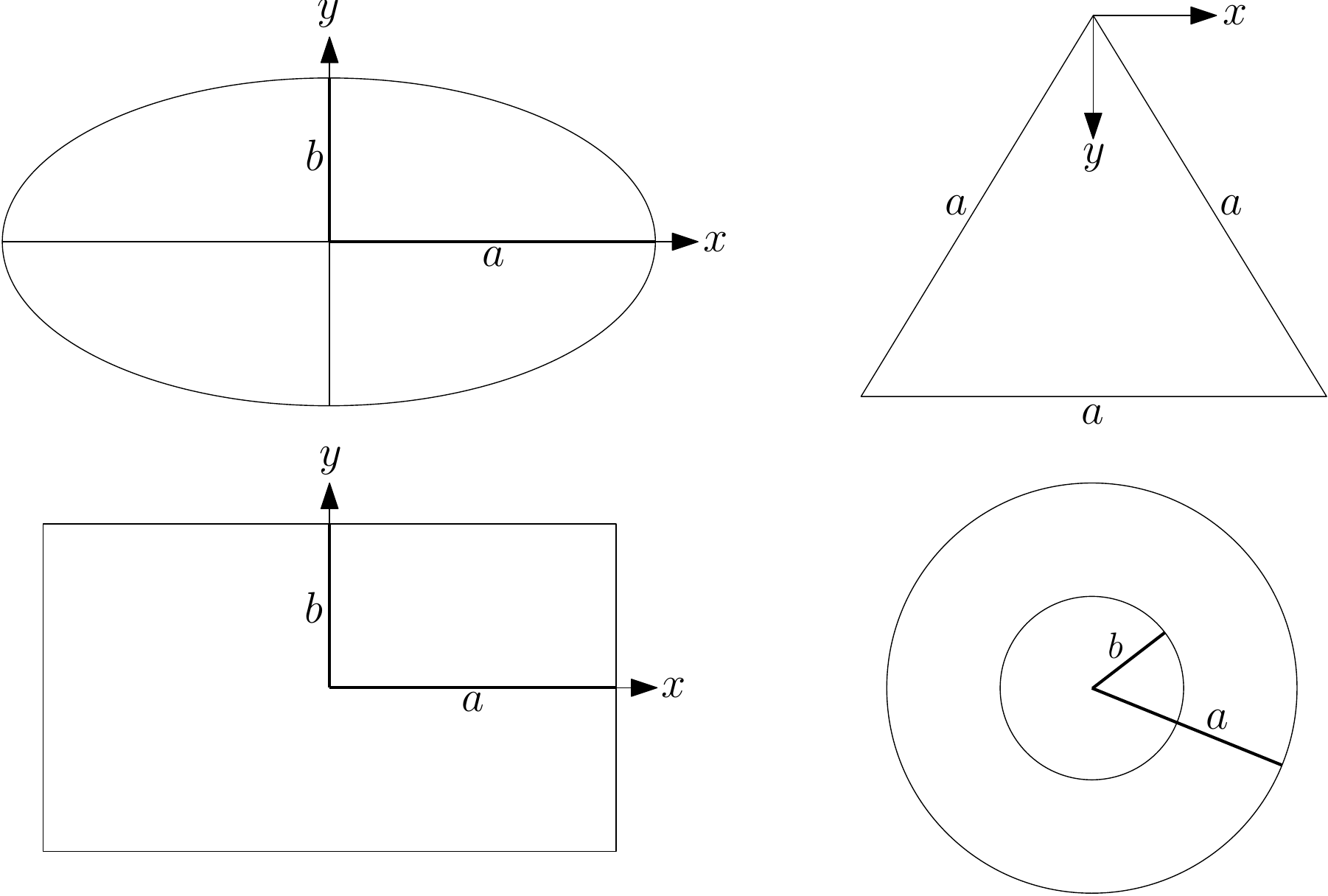} \caption{Schematics of the given examples of cross sectional shapes of
non-circular and multiply connected conduits.} \label{Shapes}
\end{figure}

For a conduit centered on the origin of coordinates with an elliptical cross section of semi-major
axis $a$ along the $x$ axis and semi-minor axis $b$ along the $y$ axis (refer to Figure
\ref{Shapes}) we have

\begin{equation}
\tau_{xz}=-\frac{\partial p}{\partial z}\frac{b^{2}x}{a^{2}+b^{2}}
\end{equation}

\begin{equation}
\tau_{yz}=-\frac{\partial p}{\partial z}\frac{a^{2}y}{a^{2}+b^{2}}
\end{equation}

For a conduit with an equilateral triangular cross section of side $a$ in the coordinates system
given in Figure \ref{Shapes} we have

\begin{equation}
\tau_{xz}=-\frac{\partial p}{\partial z}\frac{\sqrt{3}}{a}\left(\frac{a\sqrt{3}}{2}-y\right)x
\end{equation}

\begin{equation}
\tau_{yz}=-\frac{\partial p}{\partial z}\frac{1}{2\sqrt{3}a}\left(-3x^{2}+3y^{2}-a\sqrt{3}y\right)
\end{equation}

For a conduit centered on the origin of coordinates with a rectangular cross section of half length
$a$ along the $x$ axis and half width $b$ along the $y$ axis (refer to Figure \ref{Shapes}) we have

\begin{equation}
\tau_{xz}=-\frac{\partial p}{\partial z}\frac{8b}{\pi^{2}}\sum_{i=1,3,5,\ldots}^{\infty}\frac{\left(-1\right)^{\left(i-1\right)/2}}{i^{2}}\frac{\sinh\left(i\pi x/2b\right)}{\cosh\left(i\pi a/2b\right)}\cos\left(i\pi y/2b\right)
\end{equation}

\begin{equation}
\tau_{yz}=-\frac{\partial p}{\partial z}\left[y-\frac{8b}{\pi^{2}}\sum_{i=1,3,5,\ldots}^{\infty}\frac{\left(-1\right)^{\left(i-1\right)/2}}{i^{2}}\frac{\cosh\left(i\pi x/2b\right)}{\cosh\left(i\pi a/2b\right)}\sin\left(i\pi y/2b\right)\right]
\end{equation}

All these equations can be verified by substituting these expressions into Equation \ref{MomEq3}
which produces an identity in all cases.

Similarly, for a concentric circular annulus with an inner radius $b$ and an outer radius $a$
(refer to Figure \ref{Shapes}), using a cylindrical coordinates system whose $z$-axis is oriented
along the annulus axis of symmetry, we have

\begin{equation}
\tau_{rz}=-\frac{\partial p}{\partial z}\frac{1}{4}\left[2r+\frac{\left(a^{2}-b^{2}\right)}{\ln\left(b/a\right)}\frac{1}{r}\right]
\end{equation}

\begin{equation}
\tau_{\theta z}=0
\end{equation}
The latter can be verified by substitution in the $z$-component of the cylindrical form of Equation
\ref{MomEq2}, that is

\begin{equation}
\frac{1}{r}\frac{\partial\left(r\tau_{rz}\right)}{\partial r}+\frac{1}{r}\frac{\partial\tau_{\theta z}}{\partial\theta}=-\frac{\partial p}{\partial z}
\end{equation}

\clearpage
\section{Conclusions} \label{Conclusions}

In this study we propose extending the stress function approach, which was established previously
\cite{SochiStress1D2015} for the case of one-dimensional simply connected flows, to obtain
analytical or semi-analytical or highly accurate numerical solutions for the flow of generalized
Newtonian fluids in conduits with non-circular or with multiply connected cross sections. The
investigation is based on the assumption that the stress function for generalized Newtonian fluids
is the same for Newtonian and non-Newtonian rheologies.

If this assumption, which is established for the one-dimensional simply connected flow geometries
such as circular pipes and plane slits, can be established for the cases of non-circular and
multiply connected flow geometries then the stress function, which normally can be easily obtained
from the Newtonian flow case analytically or by series solutions or by other means, can be employed
in combination with the non-Newtonian rheological constitutive relations that correlate, explicitly
or implicitly, the rate of strain to the shear stress to obtain the rate of strain as a function of
the spatial coordinates and hence the flow velocity profile and the volumetric flow rate.

In previous studies we investigated the optimization of total stress \cite{SochiVariational2013,
SochiVarNonNewt2014, SochiSlitPaper2014, SochiPC2015} and the minimization of transport energy
\cite{SochiPresSA2014, SochiPresSA22014, SochiOptMulVar2014} in the fluid flow phenomena. If these
principles can be established for the flow through conduits with non-circular or with multiply
connected cross sections then this will add more support to the proposal presented in the current
study of the universality of the stress function since these principles are indifferent to the
fluid rheology.

\clearpage
\section{Nomenclature}\label{Nomenclature}

\begin{supertabular}{ll}
$a,b$                   &   conduit geometric parameters \\
$\frac{D}{Dt}$          &   material derivative \\
$\mathbf{g}$            &   gravitational acceleration vector \\
$k$                     &   viscosity coefficient in power law model \\
$m$                     &   indicial parameter in Cross model \\
$n$                     &   flow behavior index in power law and Carreau models \\
$p$                     &   pressure \\
$r$                     &   radius \\
$t$                     &   time \\
$\mathbf{v}$            &   fluid velocity vector \\
$x,y,z$                 &   spatial coordinates \\
\\
$\nabla$                &   gradient operator \\
$\alpha$                &   indicial parameter in Ellis model \\
$\gamma$                &   rate of shear strain \\
$\theta$                &   azimuthal angle in cylindrical coordinates system \\
$\lambda$               &   characteristic time constant in Carreau and Cross models \\
$\mu_{0}$               &   zero-shear viscosity in Carreau and Cross models \\
$\mu_e$                 &   low-shear viscosity in Ellis model \\
$\mu_{i}$               &   infinite-shear viscosity in Carreau and Cross models \\
$\mu_{r}$               &   characteristic viscosity in Ree-Eyring model \\
$\rho$                  &   fluid mass density \\
$\tau$                  &   shear stress \\
$\boldsymbol{\tau}$     &   extra stress tensor \\
$\tau_c$                &   characteristic shear stress in Ree-Eyring model \\
$\tau_{h}$              &   shear stress when viscosity equals $\frac{\mu_e}{2}$ in Ellis model \\
$\tau_{xz},\tau_{yz}$   &   shear stress components \\
\end{supertabular}

\clearpage
\phantomsection \addcontentsline{toc}{section}{References} %
\bibliographystyle{unsrt}

\end{document}